\documentclass[mathleft
]{an}
\usepackage{graphicx}
\usepackage{times}
\usepackage{epstopdf}

\begin{document}

\Pagespan{789}{}
\Yearpublication{2006}%
\Yearsubmission{2005}%
\Month{11}%
\Volume{999}%
\Issue{88}%

\title{Preliminary results from the STEPHI2009 campaign
on the open cluster NGC~1817}
\author{O.L. Creevey\inst{1,2}\fnmsep\thanks{Corresponding author:
  \email{orlagh@iac.es}\newline},
J.N. Fu\inst{3},
L. Fox Machado\inst{4},
M. Alvarez\inst{4},
J.A. Belmonte\inst{1,2},
\'{A}. Gy\"orffy\inst{5,6},
E. Michel\inst{7},
R. Michel\inst{4},
H. Parvianinen\inst{1},
J.H. Pe\~na\inst{8}
W.Z.Wang\inst{3},
X.H. Yang\inst{3},
\and
Y.P. Zhang\inst{3}
}
\titlerunning{STEPHI2009: asteroseismology of NGC1817}
\authorrunning{Creevey et al.}
\institute{Instituto de Astrof\'isica de Canarias, E-38200 La Laguna, Tenerife, Spain
\and
Departamento de Astrof\'isica, Universidad de La Laguna, E-38205 La Laguna, Tenerife, Spain
\and
Department of Astronomy, Beijing Normal University, 19 Xinjiekouwai street, Beijing 100875, China
\and Observatorio Astron\'omico Nacional, Instituto de Astronom\'ia, Universidad
Nacional Aut\'onoma de M\'exico, Apdo. P. 877, Ensenada BC, 22860, Mexico.
\and 
Lor\'and E\"otv\"os University, Department of Astronomy, H-1518 Budapest,
P.O. Box 32, Hungary
\and
Konkoly Observatory of the Hungarian Academy of Sciences, P.O. Box
67, HÐ1525 Budapest, Hungary
\and
Observatoire de Paris-Meudon, 5 Place Jansen, 92195 Meudon Cedex, France
\and
Instituto de Astronom\'{\i}a, Universidad Nacional Aut\'onoma de M\'exico,
Ap. P. 70-264, M\'exico, DF 04510, Mexico
}

\received{30 March 2010}
\accepted{1 April 2010}
\publonline{later}


\abstract{%
We present preliminary observational results of the multi-site STEPHI campaign on the cluster NGC~1817. The three observatories involved are San Pedro M\'artir (Mexico), Xing Long (China) and the Observatorio del Teide (Spain) --- giving an ideal combination to maximise the duty cycle. The cluster has 12 known 
$\delta$ Scuti stars and at least two detached eclipsing binary systems. This combination of characteristics is ideal for extracting information about global parameters of the targets, which will in turn impose strict constraints on the stellar models. 
From an initial comparison with stellar models using the known fundamental parameters,
and just the observed pulsation frequencies and measured effective temperatures, it appears that
a lower value of initial helium mass fraction will most likely explain the observations of these stars.}
 
\maketitle

\section{Introduction}

$\delta$ Scuti stars are 1.5 -- 2.5 M$_{\odot}$ mostly MS stars, with few observed
pulsation modes.
Interpreting these pulsations 
is difficult, mainly because
we do not know the fundamental parameters of the star.
One possibility of overcoming this difficulty is 
studying 
pulsating stars in an open cluster
where some fundamental
parameters can be determined prior to using 
any pulsation frequencies.  
In this scenario
all of the stars would be expected to have the same age and initial chemical
composition --- three of five main parameters needed to describe the stellar evolution
and structure of the star (mass and convection parameter are unknown).  
Additionally their brightness would rank the objects in terms of masses.

NGC~1817 ($\alpha = 05^{\rm h} 12^{\rm m} 12^{\rm s}$, $\delta = +16^{\circ} 41' 00"$, J2000)
is an open cluster with 12 known pulsating $\delta$ Scuti 
stars (\cite{are05}).
It has also been studied by various authors and so the fundamental parameters
are known with good precision.
One of the most recent studies confirmed it as being a metal
poor cluster with  [Fe/H] = --0.33 $\pm$ 0.08 (\cite{cp05}), 
consistent with [Fe/H] = --0.34 $\pm$ 0.26 from \cite{bn04}. 
Balaguer-Nu\~nez et al. (2004) also determined an age of 1.12$^{+0.14}_{-0.12}$ Gyr
and the distance modulus, which 
yields a distance of 1,513$^{+580}_{-380}$ pc. 
\cite{hh77} and \cite{db00} determine E(B-V) = 0.23 and 0.33 respectively.

The international STEPHI (STEllar Photometry International) network has been operative since
1987.  This network organizes ground-based photometric campaigns on 
pulsating stars, and has had many scientific successes (Michel \& Baglin 1991;
Michel et al. 2000;
Li et al. 2004; Costa et al. 2007; Fox Machado et al. 2007; Fox Machado et al. 2008).
Because some fundamental parameters of NGC~1817 have been well-determined, and
some $\delta$ Scuti stars have been confirmed, the 
2009 STEPHI campaign was dedicated to this open cluster.
In this paper we discuss some preliminary results on the ground-based 
photometric campaign which took place in December 2009.

\section{Observations and Analysis}

\begin{figure}
      \center{\includegraphics[width = 0.48\textwidth]
	{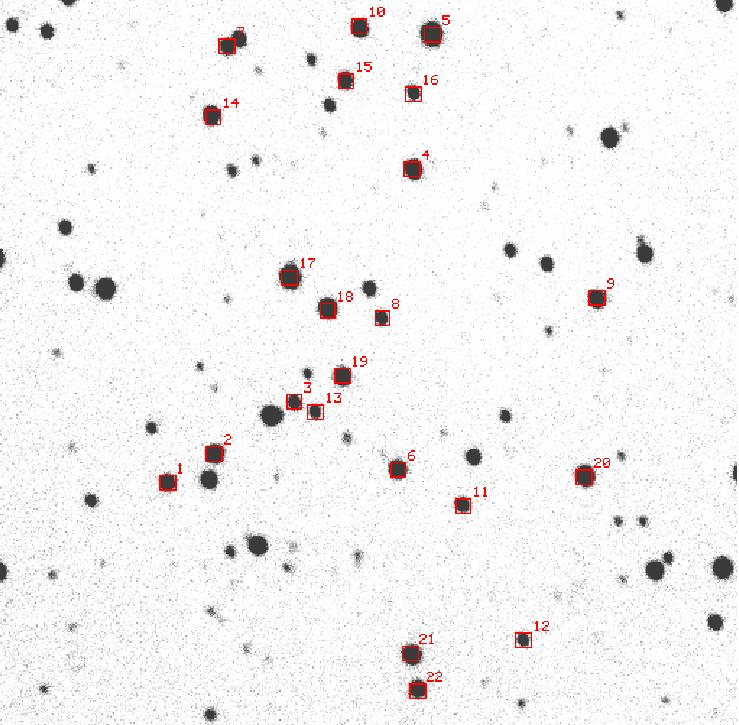}}
      \caption{The observed field of the open cluster NGC~1817 showing 
      the monitored $\delta$ Scuti stars.
	\label{fig1}}
    \end{figure}

We obtained a total of 46 nights of observations between three telescopes 
located around the globe;
the 84cm at San Pedro M\'artir in Mexico,
the IAC80 at the Observatorio del Teide, Spain, and
the 85cm at the Xinglong Station in China.
Table \ref{table1} summarizes the observing log, and
Fig.~\ref{fig1} shows the cluster field and the monitored $\delta$ Scuti stars.

\begin{table}
    \centering
      \caption{Observing Log\label{table1}}
      \begin{tabular}{llllllllll}
	\hline\hline
	Station & Telescope&Nights & Useful\\
	&& Awarded& Nights\\
	\hline
	\\
	San Pedro M\'artir (Mexico) & 84cm & 18 & 14\\
	Tenerife (Spain) & IAC80 & 14 & 4\\
          Xinglong (China) & 85cm & 14 & 10.5\\
          \\ 
	\hline
      \end{tabular}

\end{table}


    \begin{figure*}
      \center{\includegraphics[width = 0.7\textwidth]{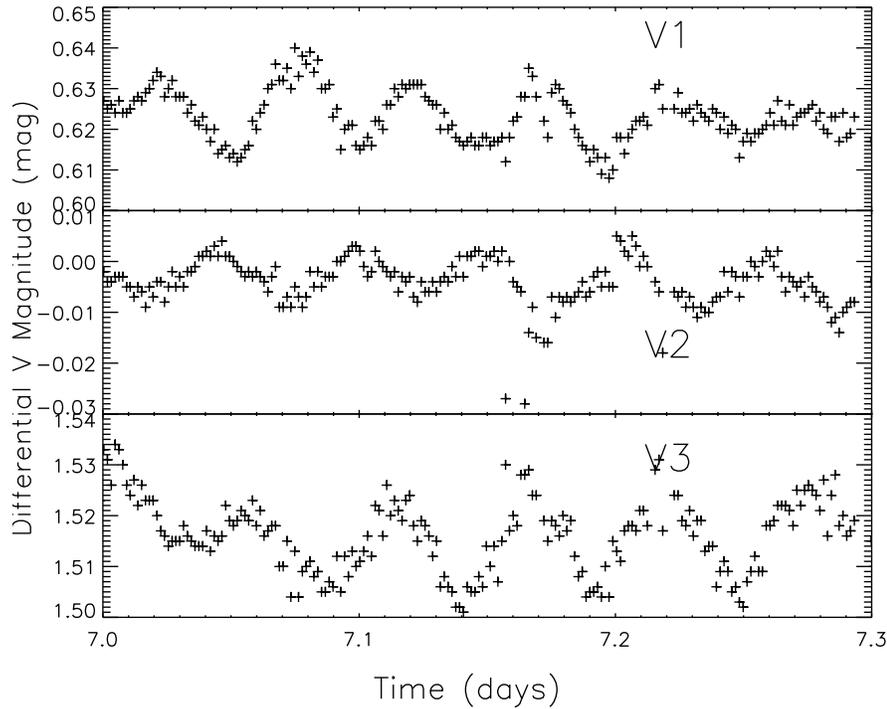}}
      \caption{Differential magnitude light curves in $V$ for 
       $\delta$ Scuti stars V1, V2, and V3, see Fig.~\ref{fig1} for reference.
	\label{fig2}}
    \end{figure*}
        

After reduction of the data, aperture photometry was performed on the candidate
stars (which include some "constant" stars) for all of the images spanning 
the 46 nights.  Light curves were then produced using differential photometry.
Figure~\ref{fig2} shows 7 hours of $V$ band light curves for the pulsating stars
V1, V2 and V3 (denoted 1 -- 3 in Fig.~\ref{fig1}).  The large amplitude pulsations of 
frequencies of 19.9, 18.6 and 18.5 cycles/day (c/d) can 
be clearly seen.

\cite{mz09} has performed an atmospheric analysis on these candidate stars, and 
they give $T_{\rm eff}$ = $7991 \pm 434$, $7298 \pm 317$, and $7962 \pm 707$ K for V1, V2
and V3 respectively using spectroscopy  and
$8118 \pm 61$, $7970 \pm 50$, and $7982 \pm 26$ K using colour information.
They also obtain $\log g$ = $3.96 \pm 0.06$, $3.50 \pm 0.06$ and $4.49 \pm 0.02$ dex.

    \begin{figure*}
      \center{\includegraphics[width=0.7\textwidth]{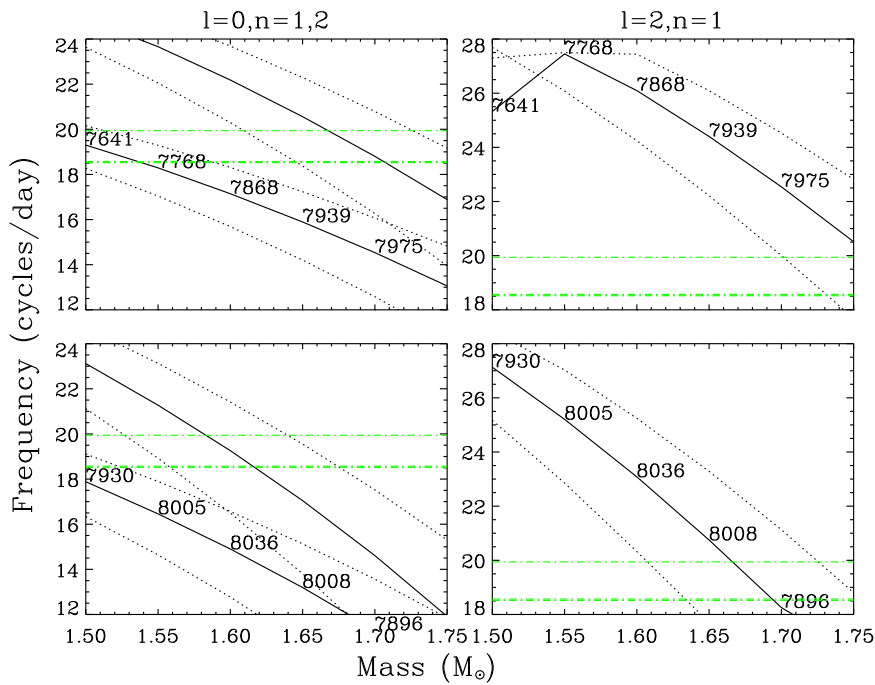}}
      \caption{
Theoretical frequencies of some oscillation modes ($l=0,2$ are left and right panels respectively) 
for stars of various masses.  
The thick line shows the frequencies for the 1.12 Gyr models and the dotted lines show the 
frequencies at the 1-$\sigma$ error bar on age.  
The numbers along the 1.12 Gyr frequency curves are the $T_{\rm eff}$ for these models.  
The green horizontal lines show the dominant pulsation frequencies 
for the three stars shown in Fig.~\ref{fig2}.  
The upper and lower panels differ mainly by the initial He mass fraction ($Y_i = 0.251, 0.282$ for 
upper and lower respectively).
	\label{fig:models}}
    \end{figure*}

\section{Initial Inference}

Due to the tight constraints on the fundamental parameters of the stars,
we can already construct some stellar models to compare the observations to.
In Fig.~\ref{fig:models} we show some theoretical pulsations frequencies 
of
low-degree ($l$), 
low radial order ($n$) modes of a 1.12 Gyr star.  
We calculated these frequencies using the ASTEC and ADIPLS
stellar structure, evolution and pulsation
codes (Christensen-Dalsgaard 2008a,b) for various masses on the main sequence.
We show the non-rotational counterparts ($m=0$) for the modes of degree
$l=0$ (left) and 2 (right).  The left panel shows the theoretical frequencies
for the radial orders $n=1$ and 2, while the right shows only the $n=1$ mode,
seen as any higher order modes would not correspond to any of the
observed pulsation frequencies (horizontal dashed lines).
The upper and lower panels show models 
for two values of initial helium mass fraction, $Y_i$ = 0.251 (top) and 0.282
(bottom).  
[$Z_i$/$X_i$] = -0.35 (top) and -0.32 (bottom), 
where $Z_i$ and $X_i$ are the initial hydrogen and 
metal content and we do not include diffusion in the models so we assume 
[$Z_i$/$X_i$] $\sim$ [Fe/H].
We show the frequencies also at the 1-$\sigma$ error bar on age and
these are denoted by the dotted lines above and below each of the solid
lines.
We have also included the theoretical $T_{\rm eff}$ of the models for 5 masses 
with an age of 1.12 Gyr,
these are the 4-digit numbers on the solid lines.

By comparing these models with just the observed pulsation frequencies and 
measured $T_{\rm eff}$,
we can hypothesise the following:
\begin{itemize}
\item If any of the stars are post MS, then they have a minimum mass of ~1.75 M$_{\odot}$.
\item Comparing initially the measured effective temperatures (if any of the stars are MS), 
it is most likely that the models with lower $Y_i$ will best represent the observations. 
\item Given the constraints on the age and metallicity we can probably disregard 
the ($l=0,n=1$) mode for the main pulsation frequency of V1 (19.9 c/d).  
This would imply a minimum mass of 1.6 M$_{\odot}$ for V1. 
\end{itemize}
Accurate apparent magnitudes and colours will constrain the masses of the star
given that they are all at the same distance.  
This in turn will allow us to constrain the identification of many modes.
    
    \section{Conclusions}
    
    We described the recent observations of the open cluster NGC~1817, and summarized its 
    known fundamental properties.  
    We performed differential photometry of the monitored stars in the cluster field and
    analyzed three of the $\delta$ Scuti stars to determine their fundamental pulsation frequencies.
    The observed pulsations are consistent with those of  Arentoft et al (2005). 
    We then constructed some stellar evolution and pulsation models for a range
    of masses and two values of initial helium mass fraction, with the 
    known fundamental properties of the cluster: Age = 1.12$^{+0.14}_{-0.12}$, and
    [Fe/H] = -0.35, -0.32, and we made some initial inferences of the three pulsating stars
    based on their observed pulsation frequencies.  
    We showed that if any of the star's are on the main sequence, then 
    it is more likely that the cluster has a lower initial helium mass fraction
    content, based on the observed pulsation frequencies and measured effective temperatures.

\acknowledgements
This research uses data obtained by the 85cm at Xinglong Observatory, China, 
the 84cm operated at the San Pedro M\'artir Observatory (SPM), Mexico 
and the  IAC80 telescope operated at the Observatorio del Teide, Tenerife. 
Special thanks are given to the night operators at Xinglong: L.G. Fang and X.Z.Yang; 
the night operators and technical staff from SPM 
and the night operators on Tenerife, 
especially Antonio Pimienta who was available to cover some nights for this campaign.  
This research uses funds from the 
Natural Science Foundation of China (NSFC) under the grant numbers 10878007 and 10778601. 
LFM and MA acknowledge the financial support from the UNAM under grant PAPIIT IN 114309. 
This research was in part supported by the 
European Helio- and Asteroseismology Network (HELAS), 
a major international collaboration funded by the European Commission's Sixth Framework Programme.


\end{document}